\documentclass[useAMS,usenatbib]{mn2e}

\usepackage{epsfig}
\usepackage{graphicx}
\usepackage{ulem}
\usepackage{color}

\DeclareMathAlphabet{\mathpzc}{OT1}{pzc}{m}{it}

\newcommand\Dl{D_{\rm L}}

\def\beq{\begin{equation}} 
\def\eeq{\end{equation}}

\def\actaa{Acta. Astronom.} %
\def\aj{AJ}%
\def\apj{ApJ}%
\def\apjl{ApJ}%
\def\aap{A\&A}%
\def\mnras{MNRAS}%
\def\pasp{PASP}%
\def\nat{Nature}%
\title[Determining neutron star masses with weak microlensing]
      {Determining neutron star masses with weak microlensing}
\author[Tian \& Mao] {Lanlan Tian$^{1}$ and Shude Mao$^{1, 2}$ \\
$^{1}$ National Astronomical Observatories, Chinese Academy of
Sciences, Beijing, 100012, China \\
$^{2}$ Jodrell Bank Centre for Astrophysics, The University of
Manchester, Alan Turing Building, Manchester M13 9PL, UK \\
}
\begin{document}
\include{journaldefs}
\date{Accepted ...... Received ...... ; in original form......   }

\pagerange{\pageref{firstpage}--\pageref{lastpage}} \pubyear{2011}
\maketitle
\label{firstpage}

\begin{abstract}
The masses of stars including stellar remnants are almost exclusively known from binary systems. 
In this work, we study gravitational microlensing of faint background
galaxies by isolated neutron stars (pulsars). We show that the resulting
surface brightness distortions can be used to determine the masses of neutron star.
Due to different evolutionary histories, isolated neutron stars may have different
masses from those in binary systems, and thus provide unique insight into their
equation of states under extreme physical conditions. We search for existing pulsar catalogues and find one
promising pair of a nearby pulsar and a background galaxy. This
method will become more practical for the next generation optical
and radio surveys and telescopes.
\end{abstract}

\maketitle


\section{Introduction}
\label{sec:introduction}

From stellar evolution, numerous ($\sim 10^8$) neutron stars are
expected in the Galaxy (e.g. \citealt{Arz02}). So far only a tiny fraction of
these have been discovered, mostly as pulsars. Even a small fraction
have measured masses, usually for binary systems (due to
additional orbital information). However, the masses of these neutron stars
may well be affected by physical processes (e.g. accretion) between the companions.
The masses of isolated stellar remnants (such as neutron stars) are thus of particular interests
since they may be more easily compared to that predicted by stellar
evolution; such comparison can provide important constraints
on the equation of state under extreme physical (e.g. density) conditions.

The recent discovery of a  $2 M_\odot$ neutron star in a binary system
already rules out some equation of state \citep{dem10}.  More mass
determinations of neutron stars in isolated or binary systems would be
important for such constraints. Gravitational microlensing has already
demonstrated its capabilities to measure the masses of isolated
normal stars (see, e.g., \citealt{Smi03, Gou09, Hwa10, Bat09}; see \citealt{Mao12} for a review).

\cite{pac95} studied how the masses of nearby dwarfs with high proper motions can be determined using microlensing,
while \cite{coles10} studied how ``weak" microlensing of background
galaxies can be used to determine the masses of nearby stars.
For the usual ``strong" microlensing,
we  observe the magnification change as a function of time; in ``weak"
microlensing, we observe the surface brightness distortions of the
background galaxy by the nearby lens as it moves across the galaxy. 
The \cite{coles10} study has one potential difficulty, that is, their
nearby stars will be very bright compared to the background galaxy,
and so some type of coronagraph has to be used to block the light from the foreground star.
In this work, we study gravitational microlensing of background galaxies by nearby stellar 
remnants ($\la 1$kpc).
The stellar remnants are expected to be faint in the optical, and so the ``glaring" problem
will be much less severe.

The outline of this paper is as follows. In \S\ref{sec:method} we first briefly recall the method of \cite{coles10}.
In \S\ref{sec:simulations}, we perform simulations for the Thirty Meter
Telescope (TMT\footnote{www.tmt.org}).  In \S\ref{sec:search} we perform a preliminary search for 
pulsars overlap with background galaxies, and present some preliminary matches. In \S\ref{sec:discussion} we discuss our results further.

\section{Method}
\label{sec:method}

In principle if we have an image of the background galaxy at the time before the lens star crosses 
in front of the galaxy and a second image when the lens star is passing the
galaxy, a simple subtraction of the two images will reveal the
distortions by the lens. From these distortions, we can derive the angular position of the lens star and the
corresponding Einstein radius. If the lens distance is known, then the  mass of the lens star can be directly estimated
from the derived Einstein radius by
\beq
\theta_E \approx  0.09''\sqrt{\frac{M/M_\odot}{\Dl/{\rm pc}}}, \label{eq:thetaE}
\eeq
where $\Dl$ is the distance to the lens, $M$ is the lens mass, and we
have assumed the source is much further away from us than the lens. 
In practice, the procedure requires the Einstein radius to be sufficiently
large and the signal-to-noise ratio sufficiently high.
	
From the presently known catalogues of neutron stars and pulsars (see section 4),
it appears rare that a galaxy is located
in the vicinity of a pulsar or a neutron star. For example, in our
search results, the smallest angular separation between
seven ROSAT discovered isolated neutron stars 
and the background galaxies is $\sim 50''$. Although these stars are nearby and have relatively high proper motions
(hundreds of milli-arcsecs per year), such a separation will still 
require a neutron star more than fifty years to overlap with the background galaxy even if it moves in
the right direction. 

We explore an alternative if we have only one image
of the background galaxy when the foreground star is already
well aligned with the background object.
If the (unlensed) surface brightness of the background galaxy is smooth on the scale of the lens distortions
($\sim 0.1''$), it is plausible that the observed distortions are due to 
the lens, from which we can derive the Einstein radius
of the lens star (see \citealt{coles10}). In practice, we may have two
images where the lens moved a detectable distance within a few years, which will allow a
difference image to be obtained; we return to this point in the discussion.

Let us consider an unlensed image $S$ 
of the background galaxy. It can be 
expanded in terms of basis functions $B^n(\vec{\theta)}$ as
\beq
S=\sum_{n} a^n B^n(\vec{\theta}),
\eeq
where $a^n$'s are the expansion coefficients. Since lensing conserves surface brightness,
the lensed image of the galaxy is
\beq
D=\sum_{n} a^n B^n(\vec{\phi}(\vec{\theta},\vec{z_t},\theta_E)),
\eeq    
where $\vec{z_t}$ is the position of the lens star and the point $\vec{\phi}$ in the source is given by 
the lens equation:
\beq
\vec{\phi}=\vec{\theta}-\theta^2_E\frac{\vec{\theta}-\vec{z_t}}{|\vec{\theta}-\vec{z_t}|^2}.
\label{eq:lens}
\eeq

Now we have a model of the lensed image $D$. Assuming the lens position is precisely known,
we have one parameter $\theta_E$, which we would like to determine,
and a set of parameters $a^n$'s, which are not of interests. Once we have an observed
 lensed image $d_{ij}$ at $t$ when the lens star has a position
 $\vec{z_t}$, the maximum likelihood is given by
 \beq
L(a^n,\theta_E)=\prod_{ij} \exp[-\frac{1}{2}\sigma^{-2}_{ij}(d_{ij}-D_{ij})^2],
\eeq 
where $\sigma_{ij}$ is the noise, $ij$ runs the pixel. We assume 
the measurement errors are Gaussian.  Because of this assumption and the fact that the model
$D$ is a linear function of the expansion coefficients $a^n$, we can marginalise the coefficients $a^n$ 
and only leave one parameter $\theta_E$ in the marginalised
likelihood (see eq. 8 in \citealt{coles10}). For clarity, we rewrite the 
marginalised likelihood as:
\beq
\chi^2=2\ln L(\theta_E)= \ln |{\rm det}\,C| + \sum_{mn}P_mP_nC_{mn} -\sum_{ij}\sigma^{-2}_{ij}d^2_{ij},
\eeq
where
\beq
P_n=\sum_{ij}\sigma^{-2}_{ij}d_{ij}L^n_{ij}
\eeq
and
\beq
C^{-1}_{mn}=\sum_{ij}\sigma^{-2}_{ij}L^m_{ij}L^n_{ij}.
\eeq
The Einstein radius $\theta_E$ is involved in the terms $L^n_{ij}$ through
\beq
L^n_{ij}=B^n(\vec{\phi}(\vec{\theta}_{ij},\vec{z_t},\theta_E)),
\eeq
which are the ``lensed" basis functions. Once the noise level $\sigma_{ij}$ is
given from the observed lensed image $d_{ij}$, we can obtain the best-fit Einstein radius
$\theta_E$ by minimising the $\chi^2$. 

With this method, even though we do not know 
the unlensed image, we can still derive the Einstein radius $\theta_E$ from
one observed lensed image. There is, however, an underlying assumption that the distortion around the lens is not due to any ``substructure" in the background galaxy.

\section{Simulations}
\label{sec:simulations}

In the section, we explore the feasibility of deriving the
Einstein radii of the pulsars or isolated neutron stars and study the
corresponding errors.

First, we produce a simulated lensed image of the background galaxy. We assume
the unlensed surface brightness follows a \cite{de48} profile.
We would like to stress that the specific form of the unlensed
image of the galaxy is not important as long as it is smooth 
on a scale a few times the Einstein radius of the lens
star. In practice, it remains to be seen whether this condition is satisfied in real images.  

The pair search in Section 4 suggests that the background source will be faint and the photons from the sky background dominate the
noise. Here we consider a specific example. We assume the source galaxy has a
magnitude $J=19.86$ and the sky surface
brightness is $J=16.75\,{\rm mag}/''^2$, appropriate for the darkest
nights on Mauna Kea \citep{San08}. It turns out
that in the region we consider ($0.284'' \times 0.284''$) the flux from the sky
background is 12.5 times that from the source galaxy. So in our simulation,
the noise comes from two parts: one from the source itself and one from the sky
background. Both follow the Poisson distribution. We then obtained the
simulated lensed image at each pixel $d_{ij}$ from this Poisson distribution.
The mean of the Poisson distribution is the 
corresponding lensed brightness $S_{ij}$ (converted into photon counts) plus photons from the sky
background.  The lensed brightness $S_{ij}$ is given by
\beq
S_{ij}(\theta_{ij})=\exp(-7.67[\sqrt{(\phi_{ij}/R_e)^2+R_c^2}]^{1/4})
\label{eq:bri}
\eeq 
with an effective radius $R_{\rm e}=1''$ and a core radius $R_{\rm c}$ of 2 pixels
(which mimics seeing). The connection between $\theta_{ij}$
and $\phi_{ij}$ in Eq. (\ref{eq:bri}) is given by Eq. (\ref{eq:lens}).
The noise $\sigma_{ij}$ is taken as
$\sqrt{d_{ij}}$. 

We specialise to the case for the next-generation extremely large optical-IR
telescopes, in particular the TMT. For the
first-light instrument, the InfraRed Imaging Spectrograph (IRIS),  the pixel
size is $0.004''$ and the diffraction-limited resolution is $0.007''$ at 1
micron. In our simulation, we use the IRIS pixel size as the minimum scale in
the choice of the scale parameter $\beta$ in the basis functions (see below). 

We assume the lenses are nearby stars in the Milky Way and the background galaxies are far away.
 For the known closest isolated neutron star, the measured distance is about
 $160 \,{\rm pc}$ \citep{Pos09}, if we assume its mass is $1.4M_{\sun}$, then 
 $\theta_E \approx 0.008''$. For the pulsar (J1238+21) which is crossing
 in front of a galaxy right now, using a mass of $1.4 M_{\sun}$ and the estimated distance
 $1.7\, {\rm kpc}$, we obtain $\theta_E \approx
 0.0025''$. Considering TMT has a resolution of $0.007''$,
 we take $\theta_E=0.005''$ and $\theta_E=0.01''$ in our simulations to produce the lensed
 images.

We use shapelets as the basis functions. As we know, a faithful shapelets decomposition depends on
four parameters: the scale parameter $\beta$, the maximum order of shapelets
order {$N_{\rm max}$} and the choice of the centroid position of the galaxy
\citep{mel2007}. The minimum and maximum sizes of features that can be
resolved by the shapelets are respectively \citep{Ref03}
$\theta_{\rm min}\approx\beta(n_{\rm total}+1)^{-1/2},
\theta_{\rm max}\approx\beta(n_{\rm total}+1)^{1/2}$,
where $n_{\rm total}$ is the sum of the maximum orders for two dimensions
($N_{x, \rm max}$ and $N_{y, \rm max}$).

In our tests, we find the scale parameter $\beta$ affects the $\chi^2$ fitting the most. 
An arbitrary choice of $\beta$ can cause a mis-estimation of the coefficients
$a_n$. Since the set of $a_n$ represents the unlensed image, a
mis-estimation of $a_n$ can lead to a mis-representation of the unlensed image.
Although $a_n$'s are marginalised out, our tests show the different choices of $\beta$ can still
affect the behaviour of the effective $\chi^2$. Therefore, it is necessary to find an optimal
$\beta$ in advance. In our tests, we found the value of optimal $\beta$ changes with the
size of the pixel and the Einstein radius. In this simulation, we use the pixel size of $0.004''$,
and $\theta_E=0.005''$.  The maximum order of the basis function is set to $N_{\rm
max}=21$ for both two dimensions ($x$ and $y$). The centre
of the image is fixed at the centre of the grid box. \cite{Ref03} suggested that
for a given $\theta_{\rm min}$ and $\theta_{\rm max}$ a good choice
for $\beta$ is $\approx(\theta_{\rm min}\theta_{\rm max})^{1/2}$.   Following this,
we choose the minimum scale as the pixel size ($0.004''$) and the maximum scale as 
the box size ($0.284''$), and the final chosen scale 
parameter is $\beta=0.03''$.

When the Einstein radius is as small as  $0.005''$, the
area affected by lensing is small (if we consider 20 times the Einstein radius, that is
$0.1''$). If the background galaxy has a small gradient in the surface
brightness on this scale, the lensing effect is too small to be
detectable. This is easily understood: lensing conserves surface brightness, and 
a uniform source has no lensing effects. This is the case when the 
lens is far away from the centre of the galaxy. Furthermore, the signal-to-noise (S/N) ratio will be low.

In contrast, near the centre of the galaxy, the surface gradient is
high, and so even if the Einstein radius of the lens star is small, the distortion
effect at this place can still be strong enough to be detectable.
In our simulation, we assume the lens star is close to the centre of the background
galaxy. We
only use a part of the lensed image which covers $71 \times 71$ pixels. With a
pixel size of $0.004''$, the area is $0.284'' \times 0.284''$. We assume that
within this area, the  number of photons from the galaxy
is $3.5 \times 10^6$ and that from the sky background is 12.5
times higher. The centre of the galaxy is at $(0,0)$, and the lens is put at
$(2.5, -2.5)$ in units of pixel. This position of the lens corresponds 
to a distance of $0.014''$.

\begin{figure}
\centerline{\includegraphics[scale=0.40]{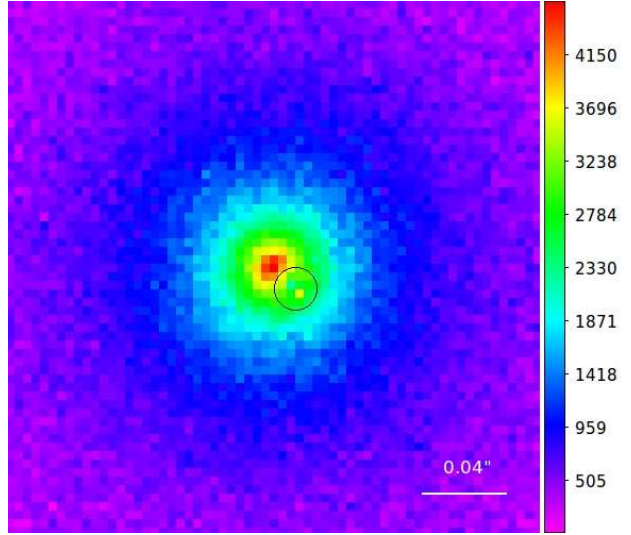}}
\caption{ The simulated lensed image by a lens star of Einstein radius
	$\theta_E=0.005''$ in the central part of the background galaxy which covers an area of $0.284'' \times 0.284''$. 
	The effective radius of the galaxy is $1''$.
	The position of the lens is the centre of the circle and the radius of
	the circle is twice of the Einstein radius. The distance between
        the lens star and the centre of the galaxy is $0.014''$. Assume the photons from sky
	background noise is 12.5 times that from the galaxy in this area. The sky
	background light has been subtracted in this image.  
}
\label{fig:lensed}
\end{figure}

\begin{figure}
\hspace{-0.5cm}
\centerline{
\includegraphics[scale=0.50]{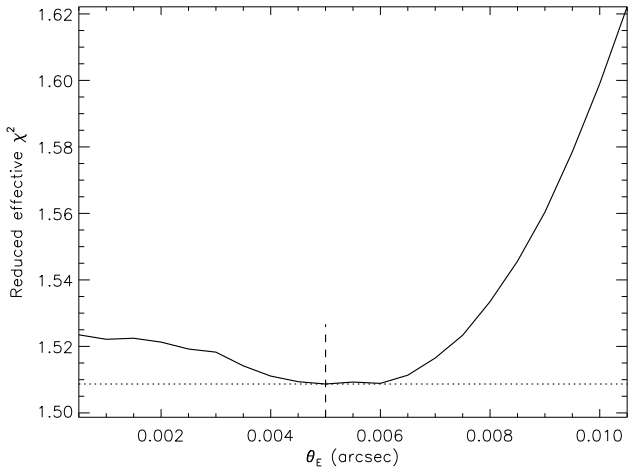}}
\caption{ The reduced effective $\chi^2$ for the case shown in Figure
\ref{fig:lensed}. The degree of freedom is 4556. The vertical dash line indicates the true Einstein radius
$\theta_E=0.005''$ used in producing the lensed image. }
\label{fig:chi_whole}
\end{figure}

The simulated lensed image is shown in Figure
\ref{fig:lensed}. The distortion within the circle in the image is caused 
by a foreground lens star with an Einstein radius $\theta_E=0.005''$. 
The centre of the circle is the position of the lens
star and the radius of the circle is $2 \theta_E$. In this case, 
the minimum $\chi^2$ value is 6873. 
The degree of freedom is
$71 \times 71-(22\times 22+1)=4556$. Then the reduced  minimum effective $\chi^2$ is
1.5. Figure \ref{fig:chi_whole} shows the reduced
effective $\chi^2 $ curve.
The best-fit $\theta_E=0.005^{+0.0003''}_{-0.0002''}$. The
$1\sigma$ error is derived from the scatters in 300 
Monte Carlo realisations. In Monte Carlo realisations,  the
simulated lensed image $d_{ij}$ are resampled from Poisson distributions.
So If there are no errors involved in
the estimates of distances of the lens star and the source galaxy, such an error
from $\theta_E$ ($\sim 6\%$) would translate into an error of $\sim 12\%$ 
in the mass estimate of the lens star ($\theta_E \propto \sqrt{M}$). 

Several factors affect the value of
the accuracy of the best-fit $\theta_E$. These are the size of the Einstein radius, the total 
number of photons and the position of the lens. For comparison, we have
repeated our Monte Carlo simulations for several other combinations of
parameters. The results are shown 
in Table \ref{table:4cases}. In this table, the maximum projected distance between the
	lens star and
the centre of the galaxy is $\sim 0.024''$. It clearly demonstrates that when the Einstein radius
increases or the lens is closer to the centre of galaxy (where the
gradient in the surface brightness is high), the Einstein
radius can be  more accurately determined. Furthermore,
when the total number of photons 
increases, the error bar decreases. 
  
\begin{table}
\caption{$1\sigma$ errors for different Einstein radii and lens positions.
	 The distance in the third row is the angular 
	 separation between the foreground lens and the centre of the background galaxy. It is in units of the Einstein radius. For each
	 simulation, the total number of
 photons from the source galaxy (in the considered region) is $3.5\times 10^6$.
 The photon number from the sky background is 12.5 times higher.
 $\theta_E$ is the true
 Einstein radius used to produce the lensed images. The $1\sigma$ errors are
 given by Monte Carlo simulations. }
\begin{tabular}{lcccl}

\hline
$\theta_E$ & $0.005''$ &  $0.005''$ & $0.01''$ & $0.01''$\\ 
\hline
lens & (5.5,-2.5) & (2.5,-2.5) &(5.5,-2.5) & (2.5,-2.5) \\ 
\hline
Distance & 4.8 & 2.8 & 2.4 & 1.4 \\ 
\hline
$1\sigma$ error & $^{+0.0004''}_{-0.0003''} $ & $^{+0.0003''}_{-0.0002''}$ & $\pm 0.00020''$
                 & $\pm 0.00015''$ \\
\hline
\end{tabular}
\begin{description}
\item[]
\end{description}
\label{table:4cases} 
\end{table}

\section{Searching existing catalogues}
\label{sec:search}

\begin{figure*}
\includegraphics[height=4.in]{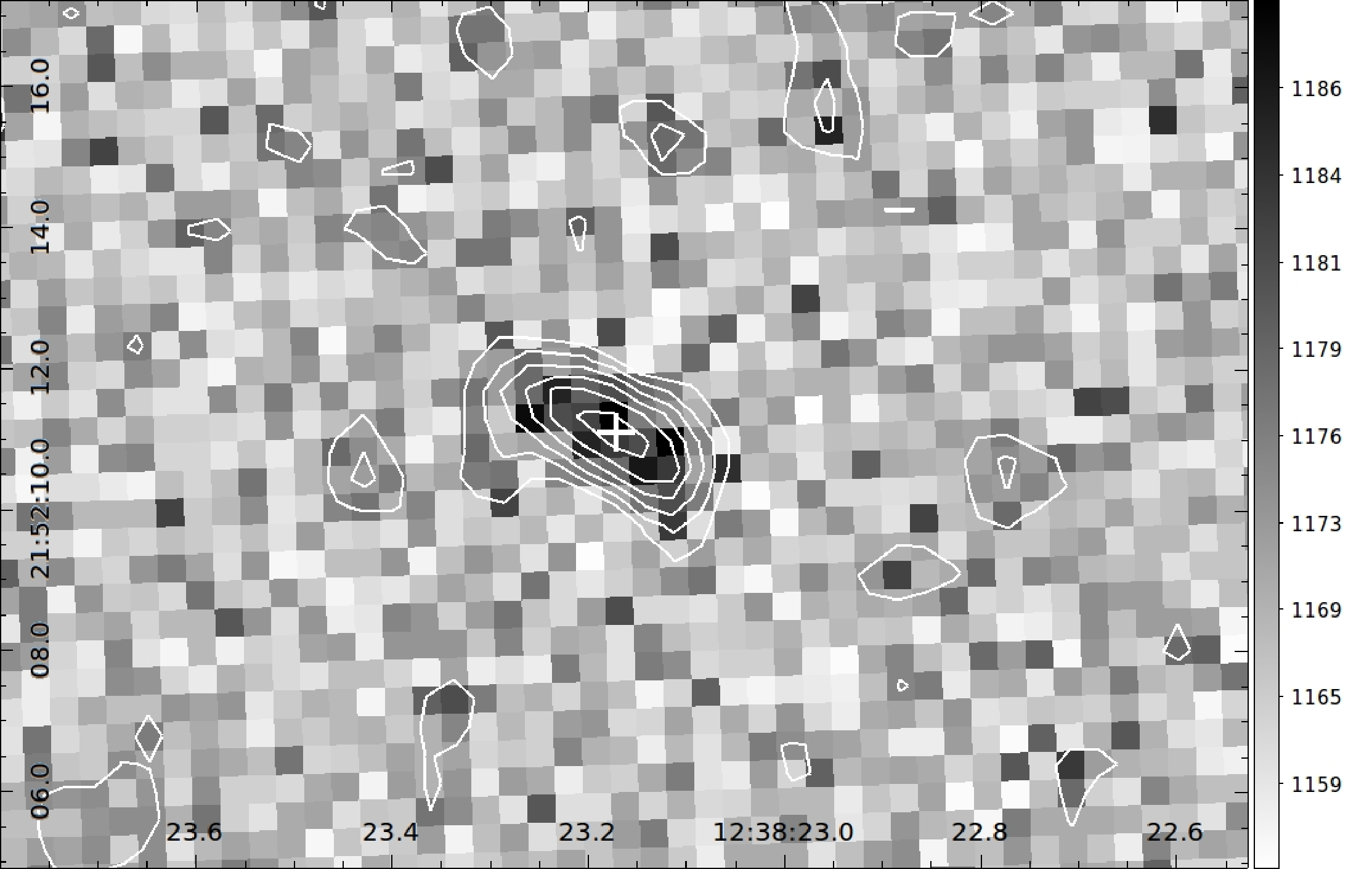}
\caption{ The SDSS image of the galaxy (Obj ID:587742013285794672). Its r-band
	magnitude is 21.45.
	At the current time,  the pulsar J1238+21 is crossing the surface of the galaxy. The
	pulsar's position in 1999 (J2000): RA
		12:38:23.17, Dec +21:52:11.1. There is no proper motion
		information available in the literature and no record shows it is a
	binary system. The white cross indicates the location of the pulsar; the error in the position is about 1 arcsec.} 
\label{fig:example1}
\end{figure*}

We use the known pulsars in the ATNF Pulsar catalogue\footnote{http://www.atnf.csiro.au/people/pulsar/psrcat/}
\citep{Man05} as our target lens
stars. We cross-match this catalogue with several galaxy catalogues, in order to look for existing or promising pulsar-galaxy
pairs. An existing pair means that the pulsar is already passing across the surface
of the background galaxy while a promising pair means that the
pulsar will move across the background galaxy within less than one hundred years. 
The data sets include
\begin{enumerate}
\item ESO archive data from NACO imaging. The search radius is $1'$.
	  We did not find any pair.
\item 2MASS Extended Source Catalog. We found 10 pulsar-galaxy pairs
	with separation of less than $5''$. Upon closer examination, we find that
	seven of these belong to the globular cluster 47 Tuc,
and two others belong to the globular cluster
NGC6544. In these cases, the globular
clusters are the ``extended'' sources rather than galaxies. Only in the
single remaining pair, the extended source is a faint galaxy.
 The information about the promising pair is listed in Table \ref{table:2mass}.
\item the SDSS catalogue \footnote{www.sdss.org}. Within a radius of $3''$, there are 4 pairs. The positions of these four
pairs are listed in Table \ref{table:pairs}. The best example is shown in Fig.
\ref{fig:example1} (the second row in Table \ref{table:pairs}). 
The other galaxies in the remaining pairs all have very low
signal-to-noise ratios. Further observations are needed to confirm whether they are true
galaxies.
\end{enumerate}

\begin{table*}
\caption{List of one pair found in 2MASS extended Source Catalog cross-matched with the catalogue of  \protect\cite{Man05}.}
\begin{tabular}{lccccccl}
\hline
Pulsar Name&  RA  & Dec & Galaxy(NED Obj Name) &  RA  & Dec & Separation & Magnitude \\
			  & (hms)&(dms)&                    & (hms)& (dms)& (arcsecs)      & (J band)\\
\hline
B1900-06 & 19:03:37.939 & -06:32:21.94 & 2MASX J19033714-0632191
			&19:03:37.1 & -06:32:19 & 3.59 & 13.161 \\
\hline
\end{tabular}
\begin{description}
\item[]
\end{description}
\label{table:2mass}
\end{table*}

\begin{table*}
\caption{List of four pairs found by cross-matching the SDSS catalog and the catalog of \protect\cite{Man05}.} 
\begin{tabular}{lccccccl}
\hline
Pulsar Name&  RA  & Dec & Galaxy(SDSS obsID) &  RA  & Dec & Separation & Magnitude\\
			  & (hms)&(dms)&                    & (hms)& (dms)& (arcsecs)    & (r band)\\
\hline
J0927+23 & 09:27:37.000 & 23:46:59.999 & 587741491435078548
			& 09:27:37.057 & 23:47:00.675 & 1.987 & 22.5 \\
J1238+21 & 12:38:23.177 & 21:52:11.075 & 587742013285794672 
			& 12:38:23.203 & 21:52:11.258 &0.414 & 21.5 \\
J2317+1439 & 23:17:09.237 & 14:39:31.219 & 587730774421078668 
			  & 23:17:09.230 & 14:39:31.182	&0.104 & 22.8 \\
B0820+02 & 08:23:09.768 & 01:59:12.412 & 588010358525657901 
			& 08:23:09.787 & 01:59:12.581 & 0.331 & 23.2 \\ 
\hline
\end{tabular}
\begin{description}
\item[]
\end{description}
\label{table:pairs} 
\end{table*}

We also searched for possible galaxies surrounding 
the seven thermally emitting radio-quiet isolated neutron stars discovered in the
ROSAT all-sky survey \citep{Motch2009}.
The galaxy search was limited within the above three databases. 
Among these, J0420.0-5022 has the closest neighbour galaxy,
$\sim 45''$ away from the star. Considering
this star has a proper motion of $123\,
{\rm mas/yr}$ \citep{Motch2009}, we have to
wait for more than 100 years for the star to move across the galaxy.

\subsection{``lensing" probability and integration time}

The above search suggests that it is rare that a known pulsar or a neutron star has a close
neighbour galaxy or is passing across a background galaxy. Since there
will be deeper sky surveys, both in the radio (the
Square Kilometer Array [SKA\footnote{www.skatelescope.org}]),
and in the optical (e.g. LSST\footnote{www.lsst.org},
PanSTARRS\footnote{http://pan-starrs.ifa.hawaii.edu/public/}), it is expected that more and
more fainter galaxies and new pulsars will be discovered in the
future. Promising pairs can then be observed with diffraction-limited, 
extremely  large optical-IR telescopes.

In this section, we will estimate, with deep imaging, the
number of galaxies that can be found in the area of sky swept out by the
known pulsars within 10 to 15 years.
From  Table \ref{table:4cases}, we see that when
the distance is $4.8 \theta_E$, we can obtain a best-fit $\theta_E$ from
one lensed image with one sigma error bar of $\sim 8\%$ 
. In the following, we
consider an area affected by a lens star is covered by a diameter of $10\theta_E$. The area swept by each
pulsar per year is $\mu\times(10\,\theta_E)$, where $\mu$ is the proper motion.   

First we estimate the total area for all the pulsars in ATNF
catalog\footnote{The catalog we used was updated on Jul 25, 2012.}.
In this catalog, there are 241 pulsars with known proper motions. Among these
pulsars, $70\%$ of them 
has a proper motion within (0, 30) ${\rm mas/yr}$. 10 ${\rm mas/yr}$ is 
the peak value of the proper motion histogram.
We then assume the remaining pulsars without 
proper motion measurements all have proper motions of 10
mas/yr. With this assumption, 
the total area swept by the pulsars in ATNF catalog within 15 years 
by $10\,\theta_E$ is 26.51  ${\rm arcs}^2$ for $\theta_E=0.01''$ and 13.25
${\rm arcs}^2$ for $\theta_E=0.005''$.
		  
We then use the GalaxyCount program\footnote{http://www.aao.gov.au/astro/GalaxyCount} \citep{Ell} to estimate the number of galaxies given
the magnitude limit.  We convert the K-band number counts to the J-band
since there is no J-band option in their program. If TMT can reach the magnitude
$24.48$ in the J-band, and we use a colour $J-K=1.0$ from UKIRT \citep{Lab10}, then we
get a magnitude limit of $K=23.48$. The program gives 115 galaxies per sq. arcmin with $K\le 23.48$. 
For $\theta_E=0.01''$, within the area, 0.85 galaxies will be found. For
$\theta_E=0.005''$, the rate is reduced by one half to about
0.42 galaxies.

It is interesting to estimate the required integration time for such a
program. 
The average number of photons is about 15 photons ${\rm s^{-1}\,m^{-2}}$ at
$J=20$ assuming a telescope throughput of about 30\%. 
In Table \ref{table:intime}, we list the integration times required for 
$\gamma_{total}=1.4\times 10^8$ for five
different magnitudes of the galaxy J=18., 19.496, 20., 20.28, 22. $\gamma_{total}$ is a 
total number of photons from the galaxy within the simulated area 
$0.284''\times 0.284''$. At faint magnitudes, the telescope time would become rather prohibitive.

We also investigate the observational feasibilities of the known pairs listed in the
tables. We assume the background galaxy has the
magnitude $19.50$ in the J-band\footnote{We use a colour $r-J=1.954$ \citep{Lab10} for the
galaxy. It may have an uncertainty of about 1 mag because we do not know the
type of this galaxy.} and an effective radius of $1''$, which
is similar to the galaxy in the second case in Table
\ref{table:pairs}. For
this magnitude, the ratio of the light from the sky background to the galaxy
itself within a central area of $0.28\times 0.28$ square arcsecs is $9.0$.
The distance between the centre of the galaxy and the lens is assumed to be $0.105''$, which is close
to the third case in Table \ref{table:pairs}. With a photon number of
$3.5\times 10^6$ from the galaxy in the given central area, the lensing effect
for an Einstein radius $\theta_E=0.005''$ is buried in the sky background.
But with a higher photon number, such as $1.4\times 10^8$, which
corresponds to a S/N $\approx 49$ at the lens position, this method can
reproduce the input value 
$\theta_E=0.005''$. However, to collect this number of photons within the given central area of the
galaxy, we need about 21 hours for TMT assuming a throughput of 30\%.

\begin{table*}
\caption{Integration time in hours needed to achieve the signal-to-noise ratio for
	TMT with 30\% throughput for $\gamma_{total}=1.4\times 10^8$. $\gamma_{total}$ is a total number of photons from
the galaxy within the simulated area $0.28''\times 0.28''$.}
\begin{tabular}{lccccl}
\hline
magnitude (J band) &  18.  & 19.496 & 20. & 20.28 & 22.\\
\hline
Integration time (hours) & 5.3 & 20.8 & 33.2 & 42.9 & 209.02 \\
\hline
\end{tabular}
\begin{description}
\item[]
\end{description}
\label{table:intime}
\end{table*}

\section{Summary and Discussion}
\label{sec:discussion}

In this work we studied the feasibility of determining the mass of isolated neutron
stars from their lensing effects on background galaxies. When a neutron
star is passing in front of a background galaxy,
the surface brightness of the background galaxy suffers from
distortions. This effect can be used to derive the Einstein radius of the lens if we {\it assume}
there are no intrinsic substructures close to the location of the lens. 

Our calculations suggest
that once the position of the 
lens star is identified precisely, the Einstein radius can be derived
from only one lensed image of the galaxy given a
sufficient number of photons. The Einstein radius can be effectively derived
when the background galaxy has a sharp surface brightness gradient near the
lens star. We showed that the Einstein radius can be derived with 4-8\%
accuracy for a size of $0.005''$ if we can achieve the required signal-to-noise ratio. The
resulting mass uncertainty is about 8-16\%. The main limitation of this
method is likely the expensive integration time required, even on extremely
large telescopes, from hours to tens of days. This long
integration time may be remedied in the future when we have larger
pulsar catalogues from SKA, or the overlapping galaxy happens
to be bright. Such examples may be already seen (see Table 3 and Fig. 3).

We are being somewhat conservative here by taking just one image. In
fact, for a relative proper motion of $10\, {\rm mas/yr}$, the pulsar would
have moved about $20\, {\rm mas}$ 
in two years, five IRIS pixels on TMT and 3 times the point spread function. One can compare the
differences in the two images, and thus obtain tighter constraints on
the lens mass. We can apply our method to the first-epoch image to infer
the lens mass, and then take a second image to further confirm and verify the model. 

If the background source is visible in the radio, then an interesting
possibility is to observe it with the Square Kilometer Array (SKA). In this
case, the resolution can be of the order of milli-arcsecs, and the dynamical
range can also be very high. However, one potential problem may be the
foreground pulsar will be bright in the radio. Similarly, in the optical,
we assumed the neutron star does not contribute any light to
the images of the lensed galaxy. However, it may not be the case in reality. 
Therefore, We need to select the optimal bands to take images of the galaxy
in order to minimise the contamination of light from the neutron star.

\section*{Acknowledgments}
We thank our referee for his/her helpful comments.
We acknowledge the Chinese Academy of Sciences for financial support and Drs. P. Saha, Charles Keeton
and Andy Gould for discussions and suggestions. LLT also would like to thank Stephen
Justham, Chen Wang and Pengfei Wang for help on pulsar astronomy. 

\bibliographystyle{mn2e}

\label{lastpage}
\end{document}